\documentclass[sn-mathphys,Numbered]{sn-jnl}
\usepackage{graphicx}%
\usepackage{multirow}%
\usepackage{amsmath,amssymb,amsfonts}%
\usepackage{amsthm}%
\usepackage{mathrsfs}%
\usepackage[title]{appendix}%
\usepackage{xcolor}%
\usepackage{textcomp}%
\usepackage{manyfoot}%
\usepackage{booktabs}%
\usepackage{algorithm}%
\usepackage{algorithmicx}%
\usepackage{algpseudocode}%
\usepackage{listings}%
\usepackage{amsmath}
\usepackage{wasysym}
\usepackage[T1]{fontenc}

\theoremstyle{thmstyleone}%

\theoremstyle{thmstyletwo}%

\theoremstyle{thmstylethree}%

\begin{document}

\title[Article Title]{LHAASO-KM2A detector simulation  using Geant4}

\author[1,2,3]{\fnm{Zhen} \sur{Cao}}
\author[4,5]{\fnm{F.} \sur{Aharonian}}
\author[6,7]{\fnm{Q.} \sur{An}}
\author[8]{\fnm{Axikegu}}
\author[1,3]{\fnm{Y.X.} \sur{Bai}}
\author[9]{\fnm{Y.W.} \sur{Bao}}
\author[10]{\fnm{D.} \sur{Bastieri}}
\author[1,2,3]{\fnm{X.J.} \sur{Bi}}
\author[1,3]{\fnm{Y.J.} \sur{Bi}}
\author[10]{\fnm{J.T.} \sur{Cai}}
% ...

\affil*[1] {\orgdiv{Key Laboratory of Particle Astrophysics \& Experimental Physics Division \& Computing Center, Institute of High Energy Physics}, \orgname{ Chinese Academy of Sciences}, \orgaddress{ \city{Beijing}, \postcode{100049}, \country{ China}}}
\affil[2] {\orgname{ University of Chinese Academy of Sciences}, \orgaddress{ \city{Beijing}, \postcode{100049},  \country{ China}}}
\affil*[3] {\orgname{ TIANFU Cosmic Ray Research Center}, \orgaddress{ \city{Chengdu},  \state{ Sichuan}, \country{  China}}}
\affil[4] {\orgname{ Dublin Institute for Advanced Studies}, \orgaddress{ \street{31 Fitzwilliam Place},\city{ 2 Dublin},  \country{ Ireland }}}
\affil[5] {\orgname{ Max-Planck-Institut for Nuclear Physics}, \orgaddress{\city{ 69029  Heidelberg}, \postcode{103980}, \country{ Germany}}}
\affil[6] {\orgdiv{State Key Laboratory of Particle Detection and Electronics}, \orgaddress{\country{ China}}}
\affil[7] {\orgname{ University of Science and Technology of China}, \orgaddress{ \city{Hefei}, \postcode{230026}, \state{ Anhui}, \country{ China}}}
\affil[8] {\orgdiv{School of Physical Science and Technology \&  School of Information Science and Technology}, \orgname{Southwest Jiaotong University}, \orgaddress{ \city{Chengdu}, \postcode{610031}, \state{ Sichuan}, \country{ China}}}
\affil[9] {\orgdiv{School of Astronomy and Space Science}, \orgname{  Nanjing University}, \orgaddress{ \city{Nanjing}, \postcode{210023}, \state{ Jiangsu}, \country{ China}}}
\affil[10] {\orgdiv{Center for Astrophysics}, \orgname{Guangzhou University}, \orgaddress{ \city{Guangzhou}, \postcode{510006}, \state{ Guangdong}, \country{ China}}}
% ...
\affil[31] {\orgdiv{National Space Science Center}, \orgname{  Chinese Academy of Sciences}, \orgaddress{ \city{Beijing}, \postcode{100190},  \country{ China}}}

\abstract{KM2A is one of the main sub-arrays of LHAASO, working on gamma ray astronomy and cosmic ray physics at energies above 10 TeV.
Detector simulation
is the important foundation for estimating detector performance and data analysis. It is a big challenge to simulate the KM2A detector in the framework of Geant4 due to the need to track numerous photons from a large number of detector units ($>$6000) with large altitude difference (30 $\rm m$) and huge coverage (1.3 $\rm km^{2}$). In this paper, the design of the KM2A simulation code G4KM2A based on Geant4 is introduced. The process of
G4KM2A is optimized mainly in memory consumption to avoid memory overflow. Some simplifications are used to significantly speed up the execution of G4KM2A. The running time is reduced by at
least 30 times compared to full detector simulation. The particle distributions and the core/angle resolution comparison between simulation and experimental data of the full KM2A array are also presented, which show good agreement.}


\maketitle

\section{Introduction}\label{sec1}

The Large High Altitude Air Shower Observatory (LHAASO) \cite{cao10} is a large hybrid extensive
air shower (EAS) array  at Haizi Mountain, 4410 m above sea level, Daocheng, Sichuan Province, China.
It is composed of three sub-arrays, the Water Cherenkov Detector Array (WCDA), one Kilometer Square Array (KM2A) and Wide Field-of-view Cherenkov Telescope Array (WFCTA) \cite{he18}.
KM2A is mainly for  gamma ray astronomy and cosmic ray physics at energies above 10 TeV \cite{liu16}.
 The whole KM2A array consists of 5216 electromagnetic detectors (EDs) and 1188 muon detectors (MDs).
The detectors were constructed since February 2018.
1/2 and 3/4 of the KM2A detectors have been operating since December 2019 and December 2020, respectively.
The whole array has been completed and operated since July 2021.

When a high energy  gamma ray or cosmic ray enters the atmosphere, it dissipates its energy
through interaction with the air molecules and causes a cascade shower. The number of secondary
particles reaching the ground depends on the energy, composition and zenith angle of the primary particles, and is also determined by the stochastic particle interaction of EAS.  EDs are capable of accurately measuring the arrival time and density of the particles of cosmic ray or  gamma ray induced shower. The time and density information are used to reconstruct the direction and energy of the primary
particle. MDs mainly detect the muons, which are used for primary component
discrimination between gamma rays and cosmic rays, or among different cosmic ray
nuclei. 
Based on the measured information, the reconstruction of primary cosmic ray information, such as the energy, composition,
flux and so on, requires the use of detector simulation techniques \cite{jchao}.

The goal of the KM2A simulation is to accurately represent the experiment.
The process of simulating extensive air showers (EAS) in the atmosphere can be effectively achieved using the CORSIKA code \cite{heck98}.
However, an accurate simulation of detector responses needs to be developed within the framework of a package such as Geant4. Each EAS event of interest contains more than 10$^6$ secondary particles that hit the KM2A array of over 6000 detectors, and possibly more than 10$^4$ photons in a single detector. Consequently, processing such a large number of photons presents a significant challenge. Meanwhile, detectors of KM2A were operated in stages during the  construction period, so a flexible, fast and accurate simulation is needed to study the performance and physics of  various geometries of the partial KM2A array. This paper introduces the KM2A detector simulation code G4KM2A, which was developed based on the Geant4 package \cite{agos03}. 

G4KM2A is an important foundation for performance study and physics analysis. It has been
successfully used for the core and angle reconstruction algorithms, serving as the basis for physics studies. Additionally, G4KM2A simulation
provides tools for studying the energy reconstruction and component selection algorithms, the $\gamma$/p discrimination algorithms.  
This simulation code has been adopted to study the performance of the half-KM2A. The angular resolution ranges from 0.5$^{\circ}$ at 20 TeV to 0.24$^{\circ}$ at 100 TeV. The energy resolution is about 24\% at 20 TeV and 13\% at 100 TeV \cite{aharon21}. The performance in gamma ray detection has been thoroughly tested using the observation of the Crab Nebula, which provides important evidence for the reliability of this KM2A detector simulation code \cite{aharon21}. Based on this simulation and pre-completed KM2A array, some exhilarating progress has been achieved by  the LHAASO collaboration, i.e.,  12 ultra-high energy (UHE, $>$ 100 TeV) gamma ray sources with the maximum energy up to 1.4 PeV were detected \cite{cao21a}. This discovery has revealed an exciting fact that the Milky Way is full of PeV particle accelerators. PeV gamma ray emission from the Crab nebula was also detected, revealing it as an extreme electron accelerator with unprecedented high accelerating efficiency \cite{cao21b}, and so on.

This paper is structured as follows: Section 2 provides a comprehensive  description of the detector simulation code, including detector construction, digital  signals and trigger logic.
The optimizations implemented to expedite runtime and minimize memory consumption of the simulation are presented in Section 3. Section 4 provides the comparison between the full KM2A data and the simulation. Finally, Section 5 provides a summary of the development of G4KM2A.

\section{KM2A Detector Simulation}
Geant4 is a toolkit for the simulation of the passage of particles through matter. Its areas of application include high energy,
 nuclear and accelerator physics, as well as studies in medical and space science. It provides a
series of tools to describe a complex geometry and material properties of an experimental setup, and the various categories of physics that handle
particle transport through detector elements. The KM2A detector simulation code is based on Geant4.

\subsection{Detector construction}
\begin{figure}[h!]
\centering

\includegraphics[width=3.0in,height=2.in]{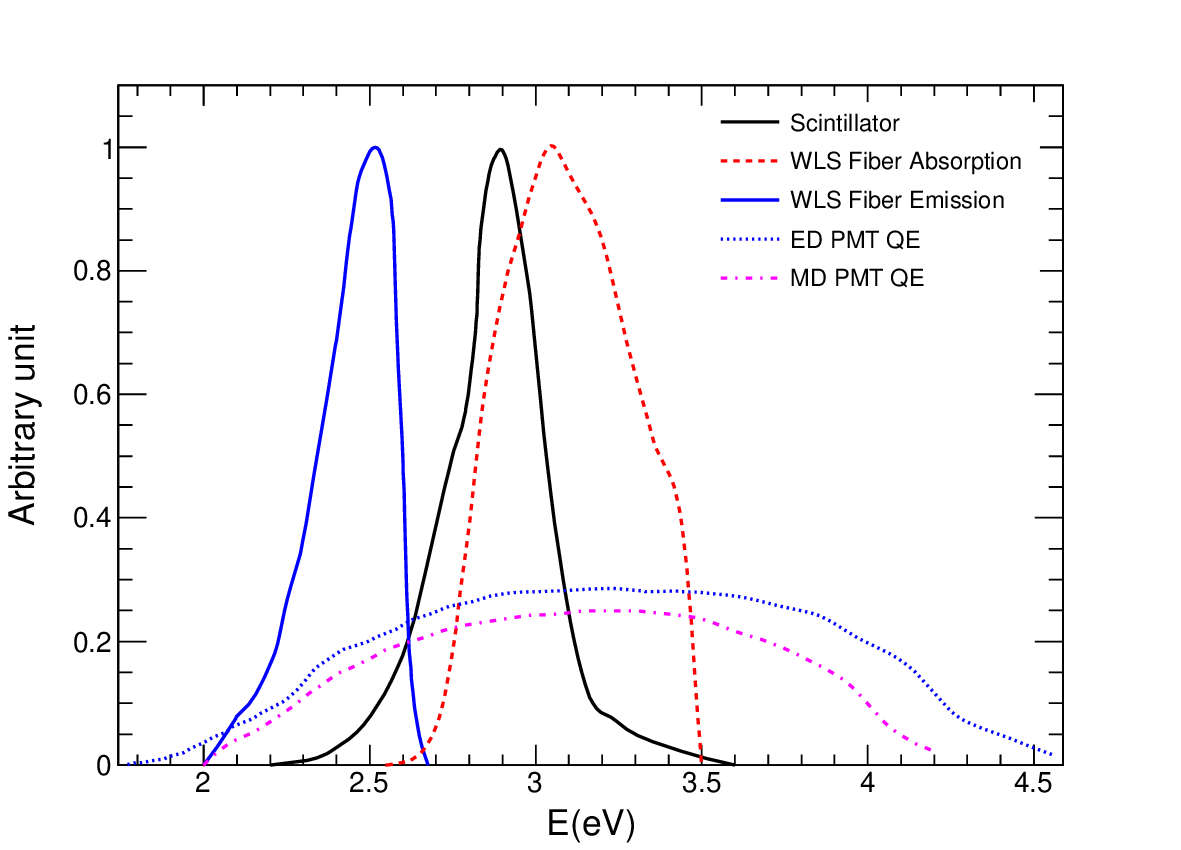}

\caption{The energy distribution of photons emitted by the scintillator (black line), absorbed by a WLS fiber (red dotted line), emitted by a WLS fiber (blue line), and the quantum efficiency of the ED PMT (blue dotted line) and MD PMT (pink dashed line). }
\label{fig1}

\end{figure}

\begin{figure}[h!]

\centering
   \begin{minipage}[t]{1\textwidth}
    \centering
      \includegraphics[width=4.5in,height=1.in,angle=0]{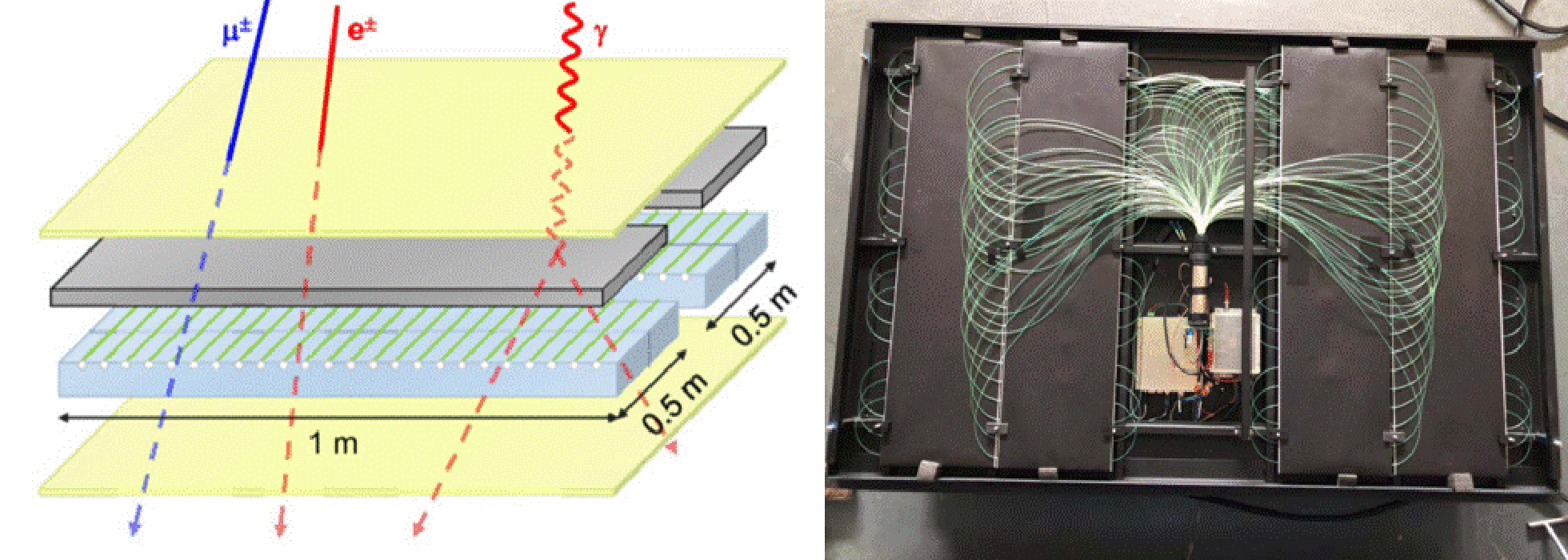}
     
    \end{minipage}

   \begin{minipage}[t]{0.45\textwidth}
    \centering
      \includegraphics[width=3.0in,height=2.in]{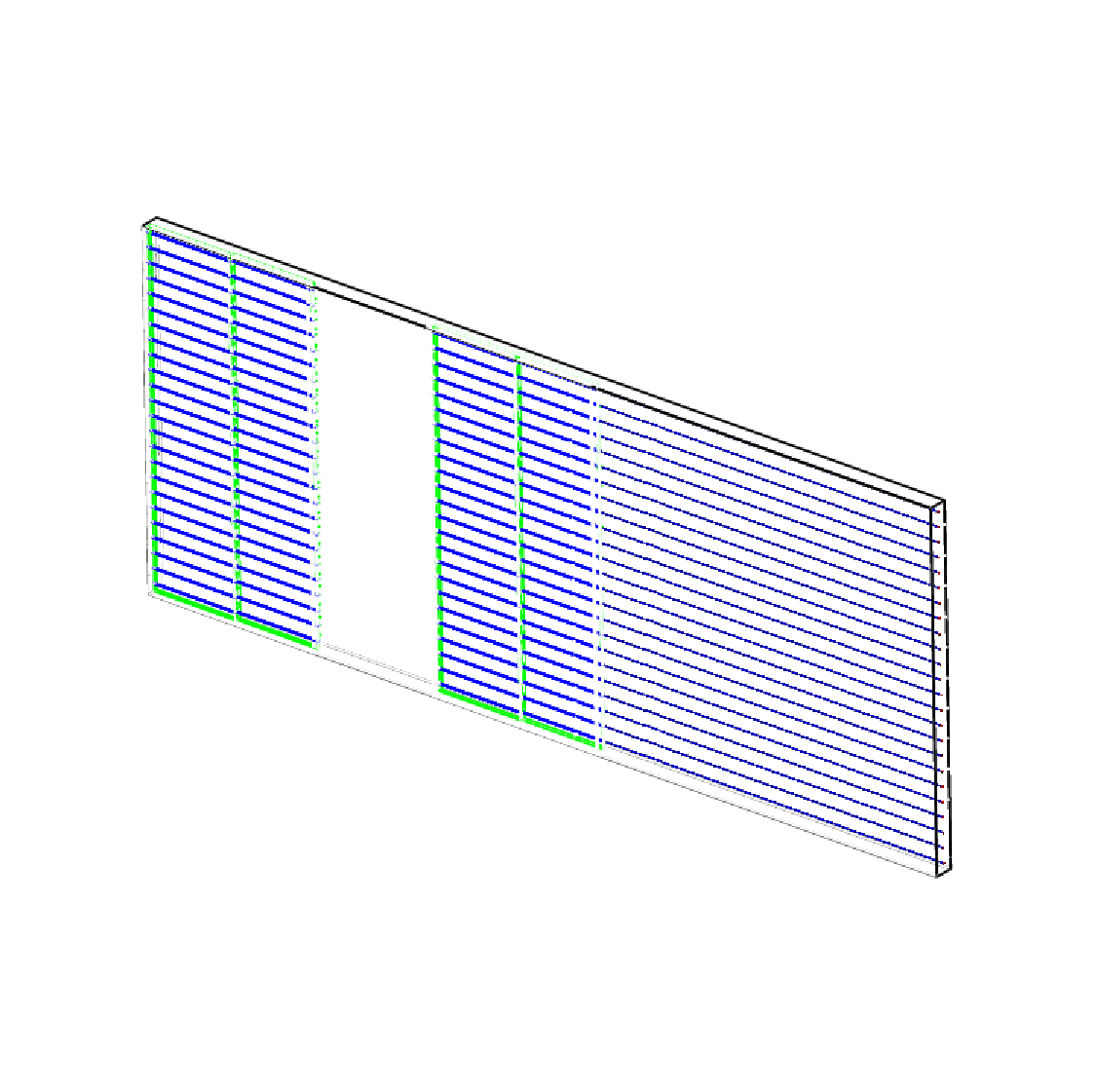}
      
  \end{minipage}
\caption{Upper left:  The schematic structure of the ED detector. The yellow part is the ED shell, the grey boxes are the lead plates, the light blue boxes are the scintillators, the green lines are the WLS fibers. Upper right: The bird view of one ED detector. Considering the size and layout of the detector, there are PMT, electronic and power-supply in the center part. Two scintillators (wrapped by black cloth) are placed at each side of the PMT. The fibers converge in a certain order at the end face of the PMT. Lower: The ED as constructed in simulation, where the green wireframe is the scintillator tile, the red dot (at the right end) is the PMT, and the blue lines are fibers. Here, only the fibers and PMTs of the far right tile are shown.} 
\label{fig1b}
\end{figure}

In the detector construction, it is necessary to set the ED and MD material properties, including geometrical elements, materials, and certain physical properties. Specific details, such as the definition of detector materials, are not elaborated in this paper. 
An ED consists of four BC408 plastic scintillator tiles (100 cm$\times$ 25 cm $\times$ 1 cm each). When a high energy
charged particle traverses the scintillator, it loses energy and excites the scintillation medium
to emit a large number of scintillation photons that follow a specific wavelength distribution,
 as illustrated in Fig. \ref{fig1}. The  number of emitted photons is proportional to the energy lost by the charged particle.
 Each tile is coated by Tyvek 1082D, which serves to reflect the photons and amplify the signal. A dielectric surface
 is utilized to treat the boundary processes of photons with the Tyvek reflector \cite{G4book}.
The surface of each ED tile has 24 evenly distributed grooves (each 1.5 mm in depth and 1.7 mm in width).
In reality, one BCF92SC wavelength-shifting fiber (WLS fiber, 2.7 m in length) is placed in every third groove, resulting in 12 WLS fibers embedded in 24 grooves per ED tile as shown in upper right panel of Fig. \ref{fig1b}. The WLS fibers  absorb the scintillation photons and emit photons with a
longer wavelength, as depicted in Fig. \ref{fig1}.
As shown in upper right panel of Fig. \ref{fig1b}, all the fiber-ends from four tiles are bundled together and  connected to a
1.5-inch photomultiplier tube (PMT) XP3960 \cite{yyh}. The portion of these WLS emitted photons that satisfy total reflection travels down the fiber to the PMT. Each photon impinging on the PMT surface is counted with a certain probability, depending on the quantum efficiency (QE) of the PMT which is wavelength dependent, as shown by blue dotted line in Fig. \ref{fig1}. However, the curvature of fibers is difficult to model by Geant4. Therefore, an equivalent method is adopted in the simulation. Each unbent fiber connects to a tiny PMT that has the same radius as fiber (shown in lower panel of Fig. \ref{fig1b}), and the signals from all the tiny PMTs in one ED are added as one signal representing the ED response. This method ensures that all the photons are collected, similar to a regular PMT.  Above the coated tile is one
5 mm thick lead plate, which absorbs low-energy charged particles and converts  gamma rays into
electron-positron pairs.

\begin{figure}[h!]
\centering

\includegraphics[width=3.5in,height=2.in,angle=0]{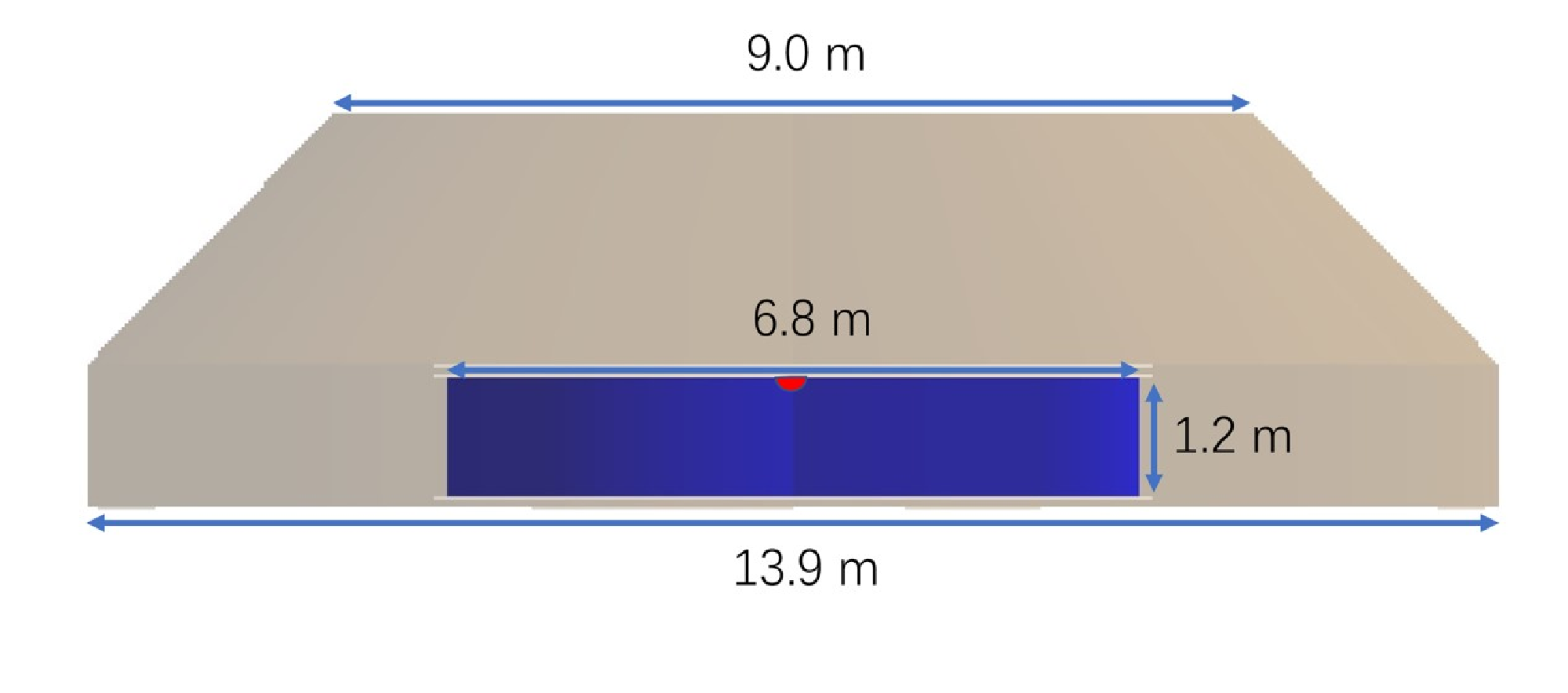}
\caption{ The geometry of the MD unit in the simulation. The light-brown region is the soil layer, and the blue box is the water bag. The red hemisphere is the PMT. The water bag is a cylinder with an inner diameter of 6.8 m and height of 1.2 m. The soil above the water bag is 2.5 m thick. The soil surrounding the tank is modeled as a cylinder with a diameter of 13.9 m.}
\label{fig2}

\end{figure}

The MD (shown in Fig. \ref{fig2}) is a pure water Cherenkov detector underneath soil. The water bag is a cylinder with an inner diameter of 6.8 m and height of 1.2 m \cite{he18}. This pure water bag is contained in a cylindrical concrete tank.
The concrete tank is buried under 2.5 m of soil, which serves to shield the electromagnetic component of the EAS.  
To simplify the simulation, only the soil surrounding the tank with a diameter of 13.9 m is constructed, as shown in Fig. \ref{fig2}. This simplification has little impact on the particles with zenith angle less than 71$^{\circ}$. It is worth to note
that such a simplification could accelerate the execution of the simulation since the transport of
high energy particles within soil is computationally intensive.
 The water bag is made of four layers of co-extruded materials,
and the inner coating is Tyvek 1082D designed to contain the Cherenkov light within the bag. The characteristics of Tyvek are simulated as for the ED. 
 When a high energy muon passes through water, it generates numerous Cherenkov photons, which are reflected on Tyvek. 
 The absorption length of light in water is wavelength dependent, and longer than 50 m \cite{he18}.
 A single 8-inch PMT (CR365-02-2 \cite{lymdpmt}) is simulated as a hemispheric volume of 10.1 cm radius immersed in the water volume
at the top center of the water bag. The PMT is defined as the sensitive detector. Each photon impinging on the
PMT surface is counted with a certain probability due to the QE of the PMT,
which is shown in Fig. \ref{fig1}.

\begin{figure}[h!]
\centering
\includegraphics[width=3.in,height=3.in]{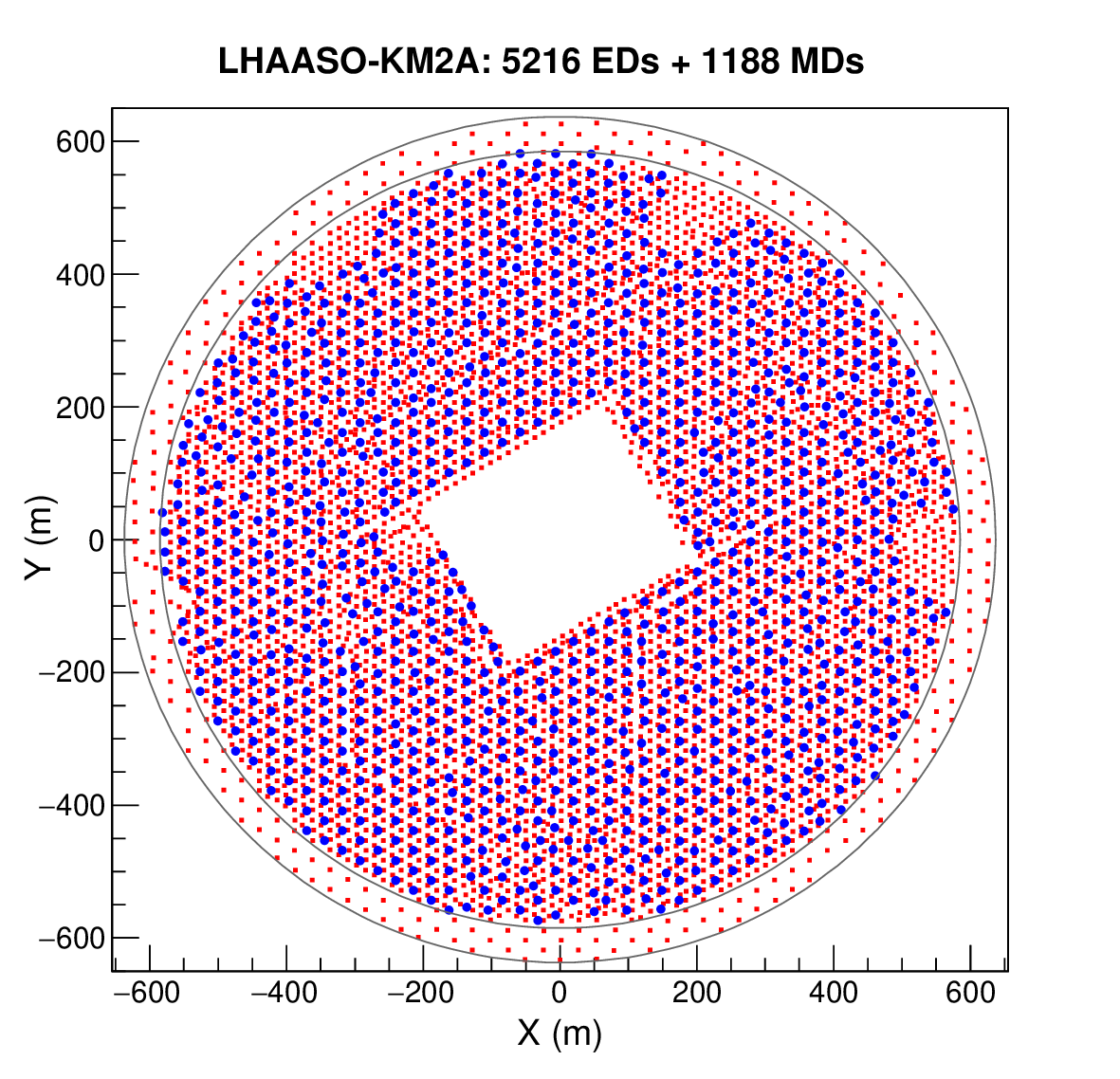}
\caption{The KM2A layout. Red dots are EDs, blue triangles are MDs. Within a radius of 575 m from the center of the array, EDs are arranged with a spacing of 15 m, while MDs are distributed with a spacing of 30 m. In the outer ring region, with a width of 60 m, the spacing of EDs is enlarged to 30 m and there are no MDs.}
\label{fig3}

\end{figure}

As depicted in  Fig. \ref{fig3}, the entire KM2A array spans an area of
1.3 km$^2$.  Within a radius of 575 m from the center of the array, EDs are arranged with a spacing of 15 m, while MDs are distributed 
with a spacing of 30 m. In the outer ring region, with a width of 60 m, the spacing of EDs is enlarged to 30 m. The outer ring is used 
to determine showers with a core located outside the array and thus improves the angular and energy resolution.
According to the geometry measurement, the maximum altitude difference of the KM2A detectors
is approximately 35 m.
 The detectors were constructed in stages, with the array being operational during construction.
To account for the real construction stages, a flexible strategy is employed in the simulation to construct the KM2A array.
 The simulation code reads an external text file containing the three-dimensional coordinates of the EDs and MDs,
and then constructs the EDs and MDs accordingly in the simulation. With the aid of this flexible detector array geometry, this simulation can be used with any configuration.

\subsection{Readout Simulation}
In the framework of Geant4, particles transport and interactions within the detector are simulated by the provided tools.
 G4KM2A tracks photons up to the PMT, after which the PMT's response and electronic readout simulation begins. The first step is to simulate the PMT response to the photons from the detector, followed by electronic readout simulation to simulate the response of electronics. In the following, we use the ED digital simulation for an example.

For a real PMT, an incident photon generates photoelectrons due to the photoelectric effect. Then the PMT converts one photoelectron into an electrical pulse. The shape of the electrical pulse reflects the performance of the PMT. In this work, the average PMT signal is obtained 
from the experimental result of single photoelectron pulses obtained by an oscilloscope. The  shape is described by the function:
\begin{equation}
\begin{aligned}
S=A\left(\frac{t-t_{0}}{\tau}\right)^{2}\ e^{-\left({ \frac{t-t_{0}}{\tau}}\right)^{2}}
\end{aligned}
\end{equation}

\begin{figure}[h!]
\centering

\includegraphics[width=3.5in,height=3.in]{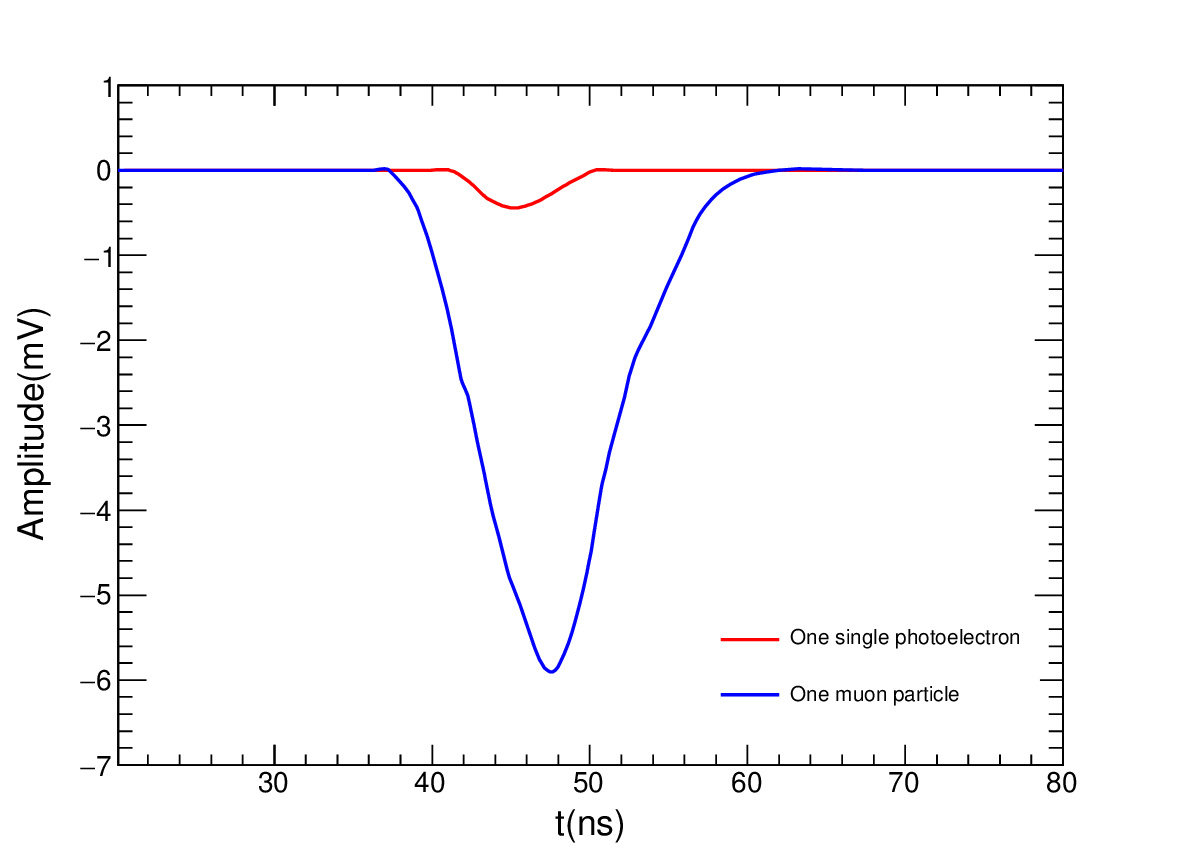}
\caption{The simulation of the ED PMT response to a single photoelectron (red line) and one muon signal (blue line).}
\label{fig5}

\end{figure}

\noindent Where \textit{A} is the amplitude, $\tau$ represents  the characteristic time of the shape, and
$t_{0}$ is the transit time in the PMT.
For each ED pulse, $\tau$ is fixed to be 3.5 ns, $t_{0}$ is sampled following a  Gaussian distribution with a mean of 30 ns and  standard deviation of 1.04 ns,
 and \textit{A} is sampled following a Gaussian distribution with a mean of 2.06 mV and a standard deviation of 0.5 mV.
For one typical minimum ionizing particle (MIP)  passing through the ED, approximately 20 photoelectrons are recorded by the PMT. The output of the signal is the superposition of these single photoelectronic pulses.   Fig. \ref{fig5} illustrates  the simulation of one ED PMT response to a single photoelectron and a muon, respectively.

Then the electronic readout simulation begins. The simulation  provides  time
information and signal information  when the PMT response's amplitude is higher than the set threshold.
The time information is the threshold-crossing time as converted  by a time-to-digital converter (TDC).
The signal information  is the charge integration as converted by an analogue-to-digital converter (ADC). Those are all based on the real electronic logic. 
Based on the characteristics of electronic devices used in the experiment, the dead times of the TDC and ADC of ED electronics  are
set at 16 ns and 400 ns, respectively. For the MD detector, the digital simulation process is the same as for the ED. The typical parameter for the MD's PMT is $\tau$=4.2 ns. The dead time  is set to be 608 ns for both TDC and ADC.

\begin{figure}[h!]
\centering

\includegraphics[width=2.5in,height=2in]{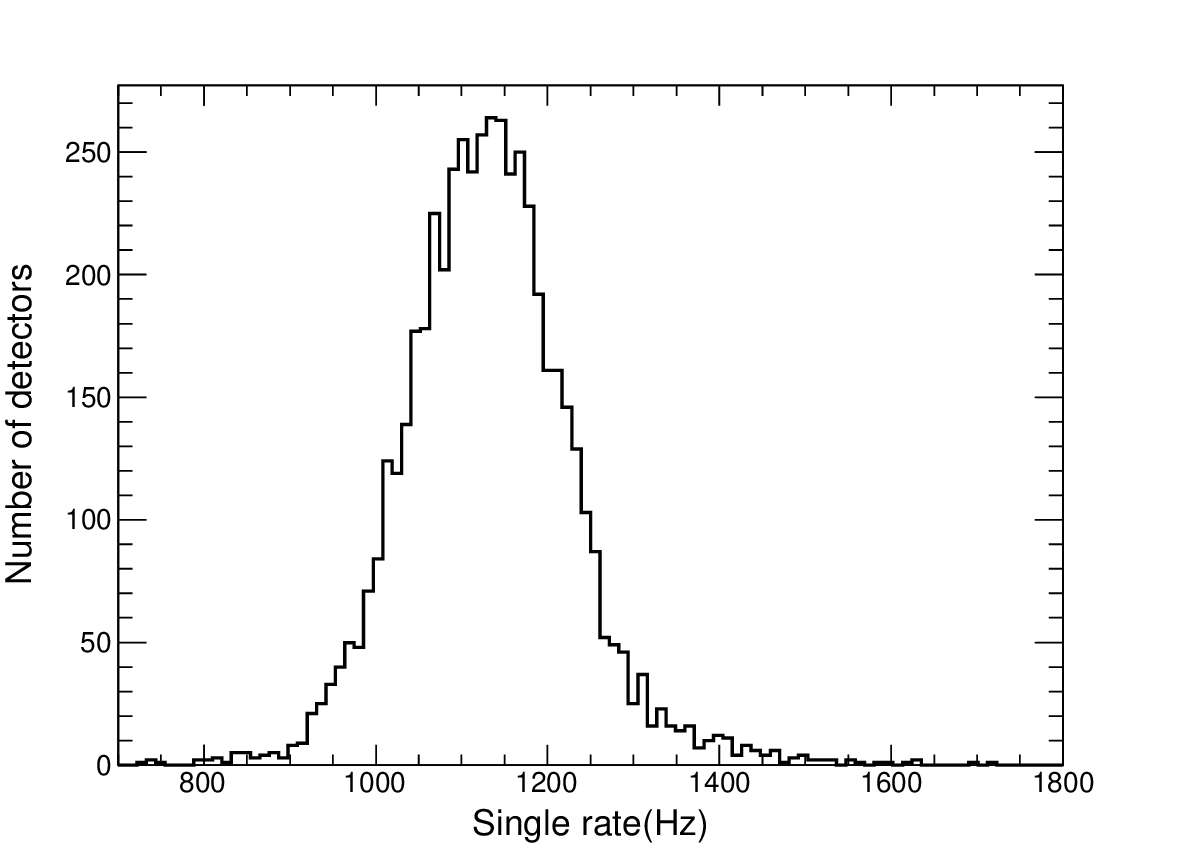}
\includegraphics[width=2.5in,height=2in]{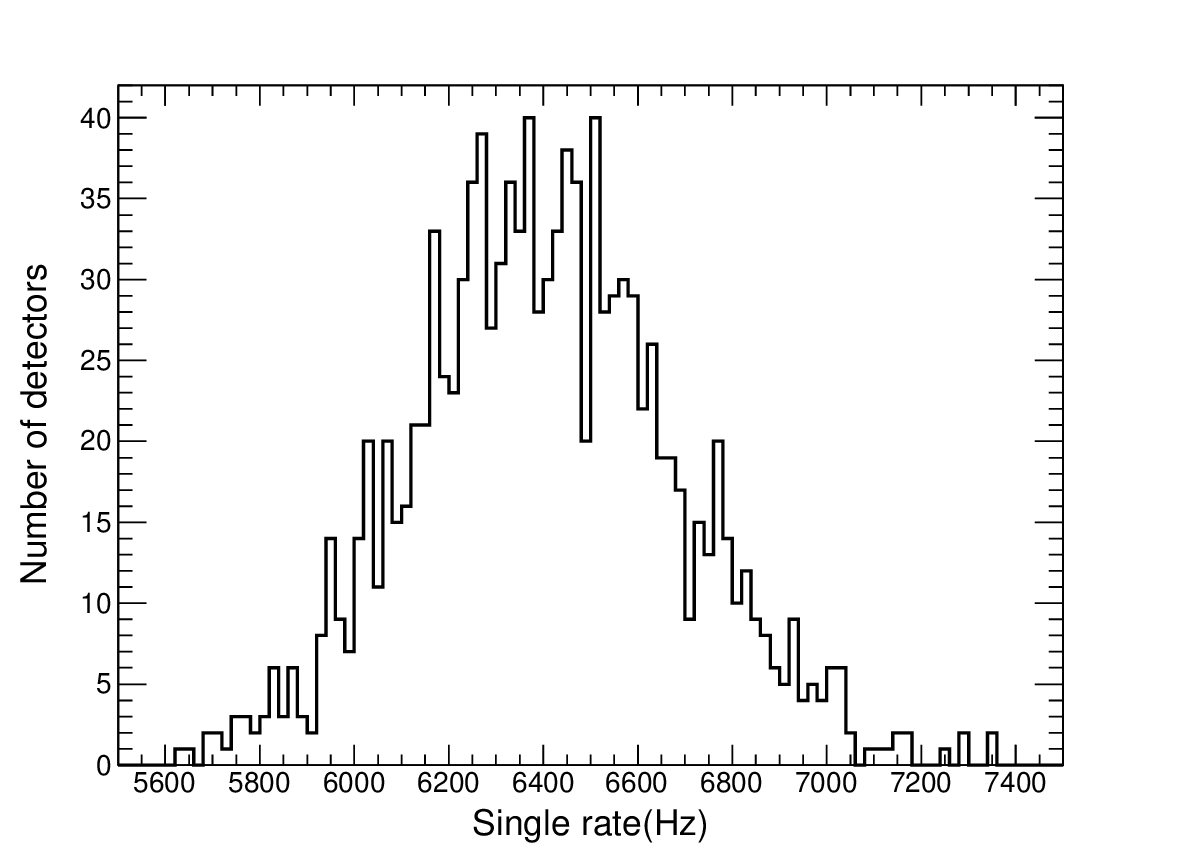}
\caption{The noise rates of EDs (left) and MDs (right) of the full array from the real data.}
\label{fig6}

\end{figure}

For each event, the recorded time window is 10 $\mu$s. During this period, some random
noise hits from single particles, ambient radioactivity, and electronic noise can also trigger the detector.
According to the data from the KM2A array at the LHAASO site,
the noise hit rate of the ED unit is about 1.19 kHz and that of the MD unit is about 6.4 kHz (as shown in Fig. \ref{fig6}),
which are adopted in this simulation. The arrival times
of noise hits are sampled with uniformly random distribution. The signal distribution of noise hits is
sampled according to the measurement result.

\subsection{Event Trigger Logic}
When a primary cosmic ray particle hits the air, it can produce  numerous secondary particles that may trigger multiple detectors in a short period of time.
Additionally, random noise events can also trigger multiple detectors coincidentally. Therefore, trigger logic is necessary to select real shower events.
In the case of the KM2A array, only hits recorded by the ED array are used for trigger discrimination. The trigger logic is critical
in determining the energy threshold, and a well-designed logic is needed to reduce the threshold as much as possible. The trigger
logic of KM2A has been extensively explored and more details could be found in \cite{wu18}.

 In the simulation code, we provide two types of trigger logic. One only uses the time window, while the other
uses both time and space windows. The trigger logic and parameter values are flexibly set by
reading the external parameter file, making it convenient to study different trigger logic scenarios.
For the full KM2A array operation, the current trigger logic requires at least 20 detectors to fire within a time windows of 400 ns.
 Once  triggered, all the hits, including shower hits and noise hits, within a window of 10  $\mu$s centered on the start time of the trigger window are recorded as an event. The simulation trigger process is the same as in the experiment. It is worth noting that in the current trigger condition, pure random noise hits still have a possibility, albeit negligible, of satisfying the trigger logic, as estimated in \cite{wu18}. However, simulating pure noise hits is beyond the scope of this simulation. In this simulation code, at least three shower hits are required to apply the trigger logic.

\section{Optimization of the simulation}
The contents presented in the previous section are standard settings within the framework of the Geant4 toolkit.
However, in the practical applications, conducting a full simulation can be extremely time-consuming. Moreover, there is a significant probability of the full KM2A simulation failing due to memory overflow. To address these two major problems, two optimizations have been developed.

\subsection{Optimization of photon tracking}
In both gamma ray and cosmic ray studies, a large number of showers need to be simulated, each consisting of over $10^{6}$  high energy secondary particles.
 Furthermore,  for each  ED or  MD, there are also more than $10^{4}$ photons within the detector when a secondary particle passes through. 
With over 6000 detectors in the full KM2A array, tracking all photons in the Geant4 simulation becomes  extremely time-consuming, rendering it impractical for use.
Additionally, the extensive tracking process can  easily lead to memory overflows.
Therefore, it is necessary to devise specialized methods to simplify the photon simulation procedure and accelerate the simulation process.

\begin{figure}[h!]
\centering

\includegraphics[width=2.5in,height=2in]{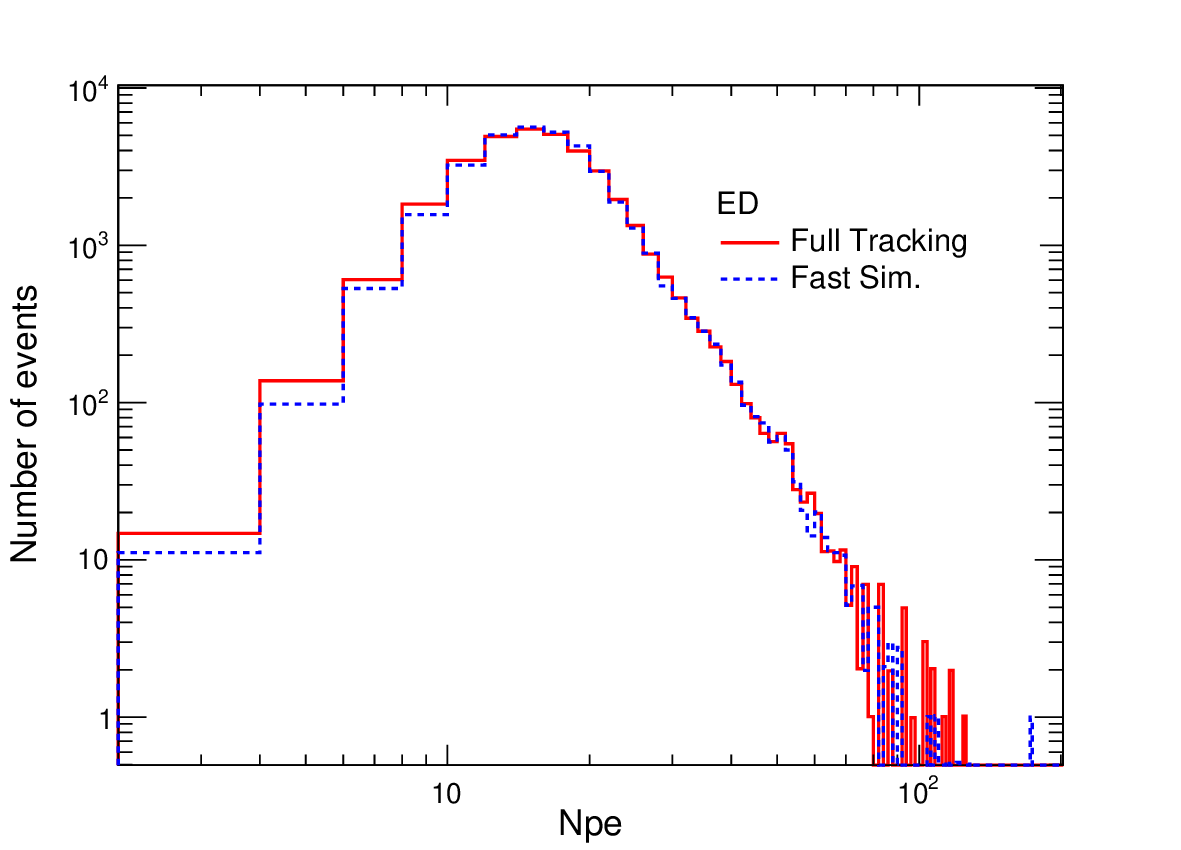}
\includegraphics[width=2.5in,height=2in]{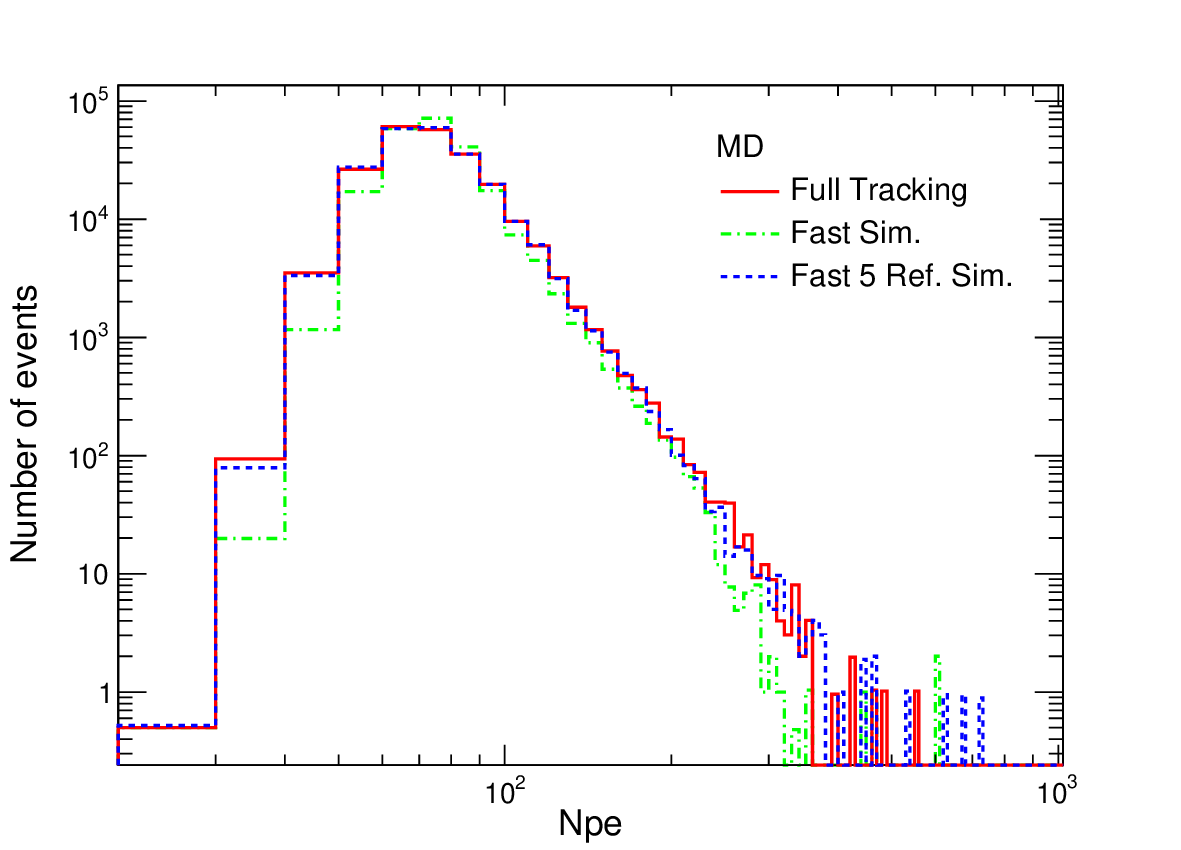}

\caption{left: The number of photoelectrons recorded by simulated ED PMT using full tracking (red line) and fast simulation (blue dotted line) method. Right:  The number of photoelectrons recorded by MD PMT using full tracking (red line), fast simulation (green dashed line), fast 5-reflection simulation (blue dotted line), respectively. }
\label{fig71}
\end{figure}
In practice, both EDs and MDs utilize PMTs as the photosensitive device. However, due to the QE of a PMT, only $\sim$ 20$\%$ of impinging photons can be recorded, as shown in Fig. \ref{fig1}. To minimize tracking of numerous
 photons, Cherenkov photons produced in the MD are randomly killed based on the QE
prior to tracking. For scintillation photons produced in the ED, fast simulation is achieved through two steps.
First, scintillation photons are randomly killed by using a constant ratio of the maximum value of the QE.
Then, the residual wavelength-dependent QE efficiency is subsequently applied during the tracking process in the WLS fiber after the wavelength shifting process.
 By implementing this optimization,  the simulation process can be accelerated by a factor of approximately 5.

\begin{figure}[h!]
\centering

\includegraphics[width=2.5in,height=2.in]{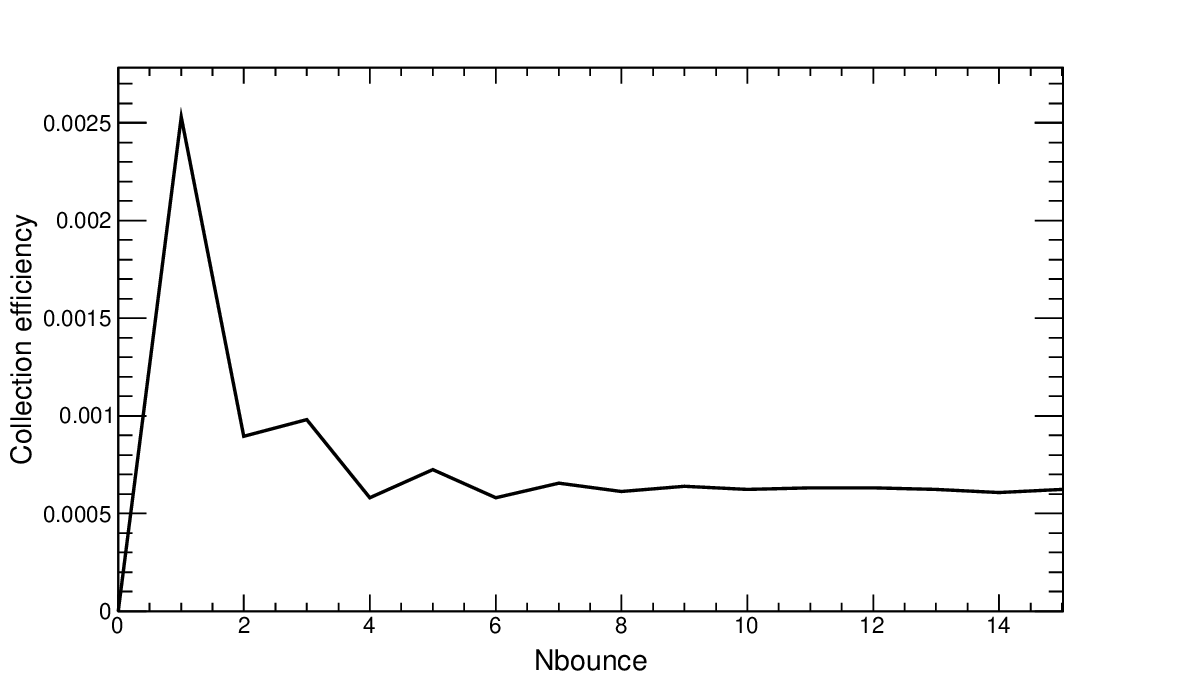}
\caption{The photon collection efficiency of an MD PMT vs. the number of reflections. N$_{\rm bounce}$ is the number of times that a photon reflects. }
\label{fig8}
\end{figure}

To further improve calculating efficiency, an optional speed-up method is also provided. Based on the tracking results of the full simulation, only a small fraction ($<$1$\%$) of photons produced in an ED or MD can propagate
to the PMT and then generate photoelectrons. In the case of the ED, where fibers are uniformly distributed in each  scintillator tile, the probability of photons reaching the PMT is
minimally dependent on position and direction. As such, after the EAS particles pass through the lead plate and generate scintillation photons in the tile, all scintillation photons' tracks
within the ED are immediately terminated, and  the number of photons that reach the PMT and their transport time are sampled from the full tracking result. Fig. \ref{fig71} illustrates the comparison
between full tracking results and fast simulation results for the number of photoelectrons recorded by PMT in
5 GeV muon simulations. The fast simulation results are consistent with full tracking simulation results, and this simplification
 can accelerate the ED simulation process by a factor of approximately 100. However, for the
MD, which has only one PMT within a volume of 44 m$^3$, and the photons are reflected by the tyvek, the probability of the photons
reaching the PMT is position and direction dependent. A  direct simplification, as used in the case of ED, would be problematic as shown by the green dashed line in the right  panel of Fig. \ref{fig71}. Fortunately, the photon collection efficiency of the PMT is almost independent on the positions and directions of photons after the fifth reflection. So in the MD simulation, photons are only tracked until the fifth reflection. After that, based on Fig. \ref{fig8}, photons are randomly killed using the collection efficiency. This simplification can accelerate the MD simulation process by a factor about 4 and remains consistent with the full tracking simulation as shown by the blue dotted line in the right panel of Fig. \ref{fig71}. Based on this optimization, the entire simulation can be sped up 30 times.

\subsection{Optimization of memory consumption}
\begin{figure*}
\centering
\includegraphics[width=6in,height=2.in]{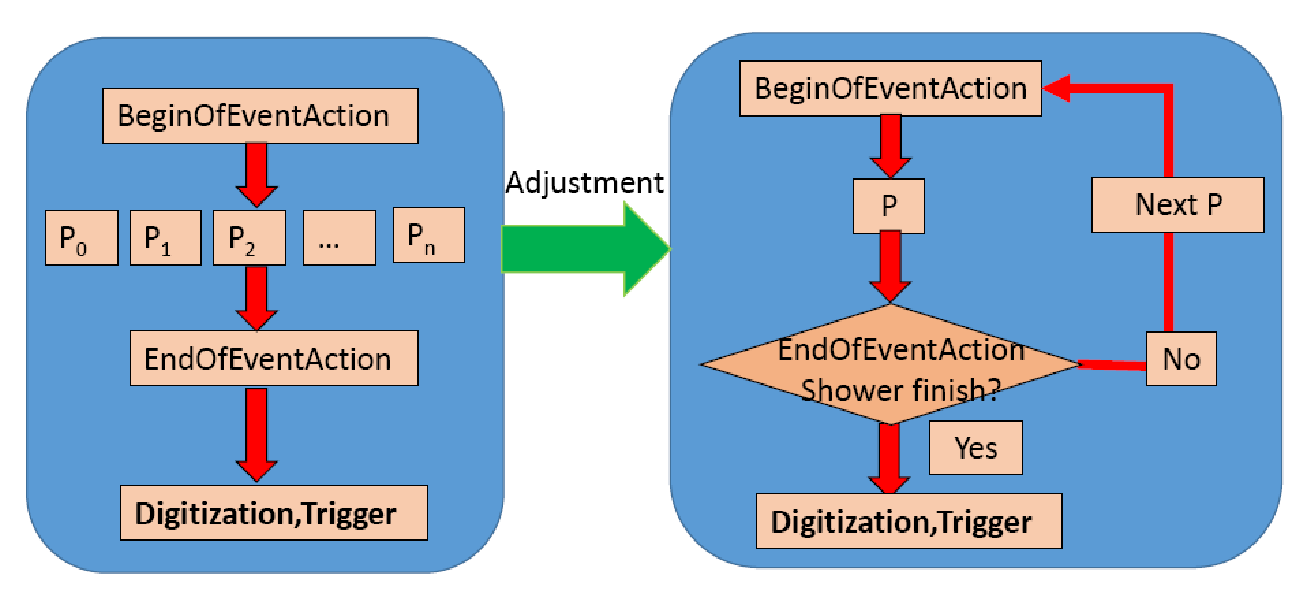}
\caption{The flow diagram of simulation process before and after adjustment. P stands for one  particle of a shower. }
\label{fig9}
\vspace*{0.5cm}
\end{figure*}

Within the normal Geant4 framework, the particle generator generates all the particles for an event, and the
internal mechanism tracks all particles and their secondary particles. During this process, computational memory is consumed to record particles and  tracking information. In the normal Geant4 simulation of an EAS event, the particle generator reads in all the particles of one EAS event produced by CORSIKA. Each EAS particle can induce a significant number of secondary particles and photons when entering an ED or MD. In such a situation, the simulation process often  crashes due to memory overflow, which requires much more than 100 GB of memory.  Even after adopting the fast simulation method mentioned in the previous section, the computational memory size limitation may still cause memory overflow. To address this issue, the simulation process flow is adjusted by reading in  the EAS particles one by one in the particle generator, instead of reading in all the EAS particles at once. For each EAS particle, after the end of the event, an internal mechanism  will store photon hit information and release the memory at this step. When the last EAS particle finishes the event action process of Geant4, the digital and trigger logic process begins to handle the photon hit information of detectors. With this optimization, the simulation process can run normally with a computational memory larger than 4 GB. The Fig. \ref{fig9} illustrates the simulation process before and after the adjustment.

It is worth noting that the impact of this adjustment on the speed of simulation execution is 
negligible compared with the detector simulation process. This optimization serves as a good example for future larger EAS experiments, which may also consist of thousands of detectors and need to process cosmic ray showers with higher energy and more secondary particles. 

\section{Simulation of the full KM2A array}
The full KM2A array has been operational since July 2021. The trigger logic requires at least 20 EDs to fire within a time window of 400 ns. The trigger rate varies between  2300 Hz and 2800 Hz due to the variation of surrounding meteorological conditions. To simulate KM2A response  to the cosmic rays, CORSIKA 7.6400  with the hadronic interaction model GHEISHA at lower energy
and QGSJET II-04 at higher energy is used to simulate the EAS process within the atmosphere.  Five groups of dominant component elements, H, He, CNO, Mg-Si, and Fe, are generated  with  energies ranging from 1 TeV to 10 PeV, zenith angles from 0$^{\circ}$ to 70$^{\circ}$, and core positions within a circular area with a radius of 1000 m centered on the array center. The fast simulation version of G4KM2A  presented in this paper is used to simulate the detector response.

\begin{figure}[h!]
\centering

\includegraphics[width=3in,height=2.5in]{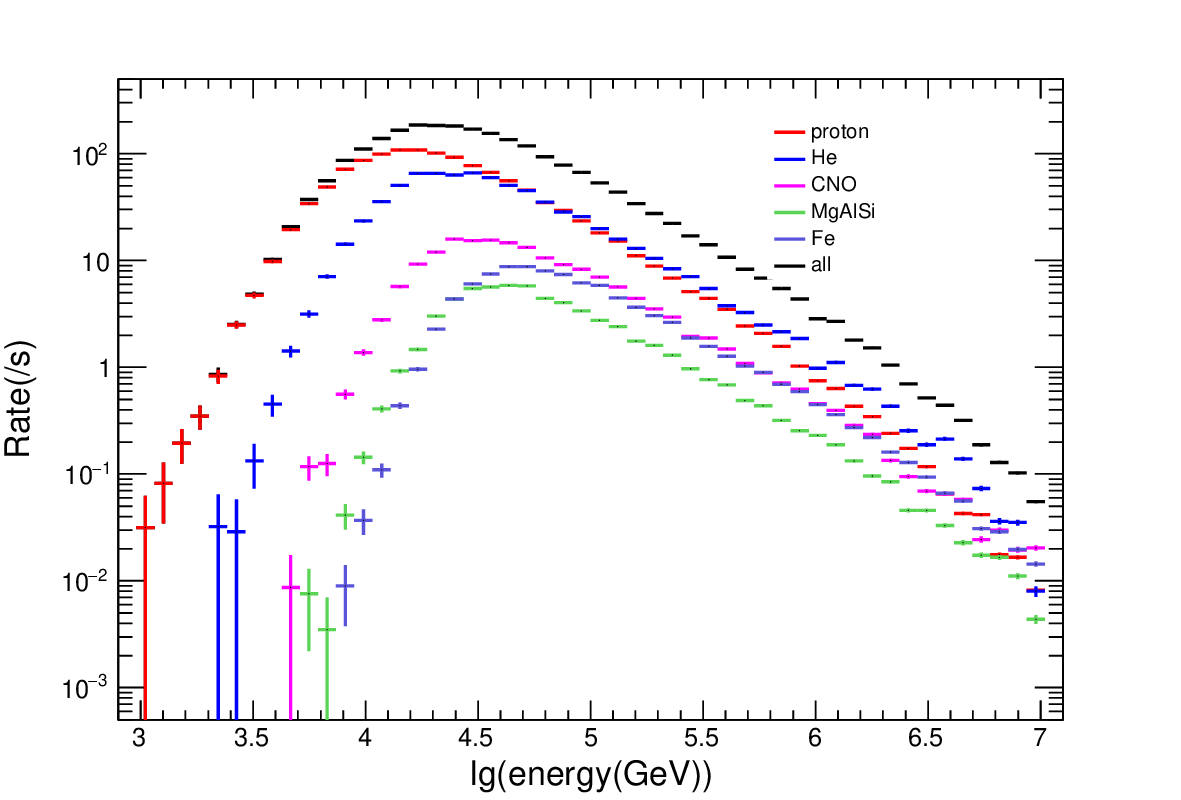}
\caption{ The energy distribution of the triggered simulation event. The component flux is normalized to H3a model.}
\label{fig10}
\end{figure}

According to the simulation, the expected  trigger rate mainly depends on the cosmic ray species and energy. 
 For the spectra of the five components, the  H3a model presented in \cite{gaiss12} is used. It is worth noting that this model may deviate from cosmic ray spectrum measurements by a few percent, based on inconsistent energy scales of different measurements achieved by different experiments. 
  The  trigger rate based on the simulation is 2265.20 Hz, including 1206.64 Hz by H, 739.61 Hz by He, 168.84  Hz by CNO, 59.76Hz by Mg-Si, 90.35Hz by Fe. These trigger rates are roughly consistent with the data. The energy distributions for the five components after trigger are shown in Fig. \ref{fig10}. The energy threshold is about 13 TeV. 

Detector simulation is an important foundation for detector performance study and physics analysis. Therefore, the consistency between simulation and experimental data is crucial.
For each recorded event, the reconstructed direction, core position and number of particles are used for scientific analysis. Therefore, the following comparison is mainly on these items. For comparison with simulation, we used experiment data collected on December 28th, 2022 and May 10th, 2022, which were randomly chosen, with trigger rates of 2658.86 Hz and 2443.26 Hz, respectively.
Both the simulation and experimental data are reconstructed by the same reconstruction process \cite{aharon21}. In the following comparison, events are selected to have a shower core inside the main KM2A area, a zenith angle less than 50$^{\circ}$ and the number of fired EDs after filtering out the noise ($N_{fit}$) larger than 10.

\begin{figure}[h!]
\centering
\includegraphics[width=2.5in,height=2.in]{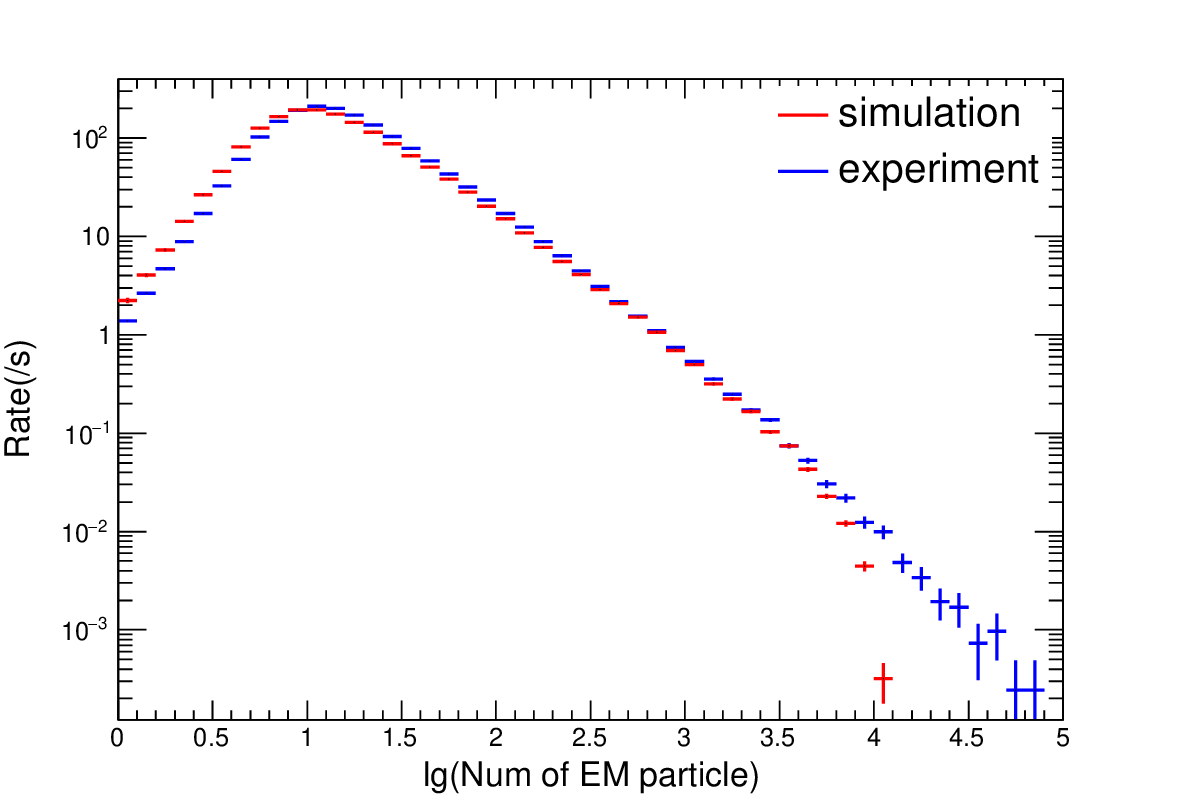}
\includegraphics[width=2.5in,height=2.in]{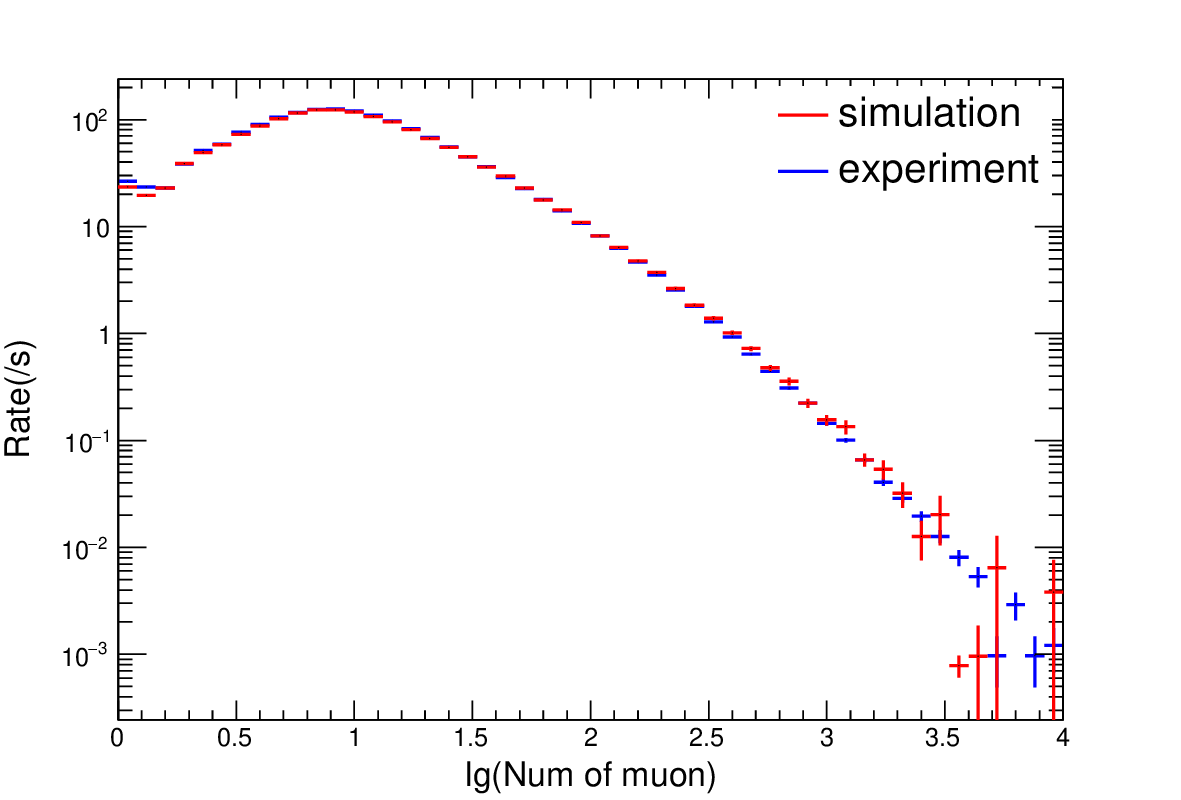}

\caption{Left: The distribution of electromagnetic particle number. Right: The distribution of muon number.  The number of electromagnetic particles ($N_{e}$) is the sum of particles recorded by EDs  within 40-100 m  from the shower core. The number of muons ($N_{\mu}$) is the sum of  muons recorded by MDs  within 15-200 m from the shower core. These distributions show that the simulation is roughly consistent with the data over four orders of magnitude.}
\label{fig11}
\end{figure}

The numbers of muons  and electromagnetic particles  are key parameters for KM2A to reconstruct the primary particle energy and to discriminate gamma rays from the cosmic ray backgrounds.
The comparison of these numbers is shown  in Fig. \ref{fig11}.  The number of electromagnetic particles ($N_{e}$) is the sum of particles recorded by EDs  within 40-100 m  from the shower core. The number of muons($N_{\mu}$) is the sum of  muons recorded by MDs  within 15-200 m from the shower core.  The simulation is roughly consistent with the experimental data within four orders of magnitude. The gap at the highest $N_{e}$ and $N_{\mu}$ values is caused by the lack of $>$10 PeV cosmic rays in the simulation. The consistency between experiment and simulation is also supported by the measurement of the spectrum of the  gamma ray standard candle Crab Nebula, which is consistent with previous measurements \cite{aharon21}.

\begin{figure}[h!]
\centering
\includegraphics[width=2.5in,height=2.in]{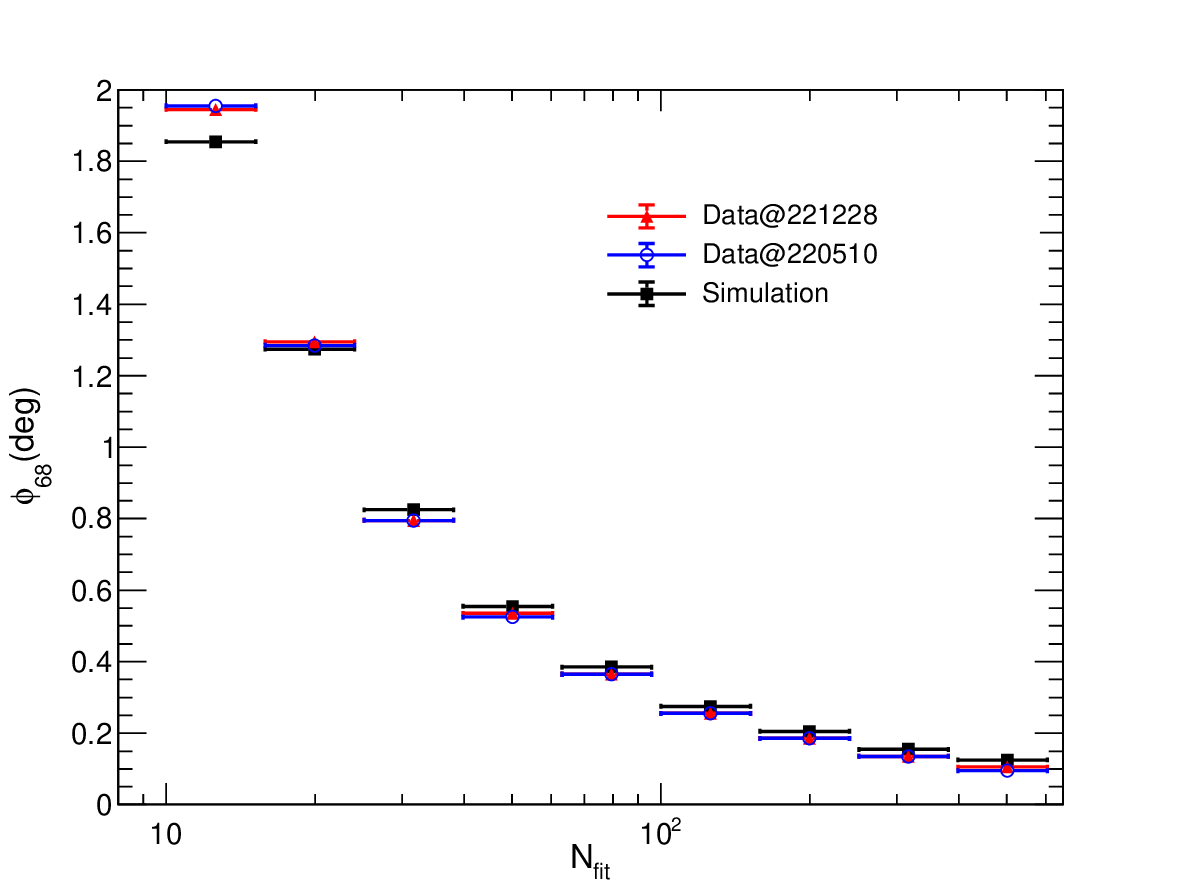}
\includegraphics[width=2.5in,height=2.in]{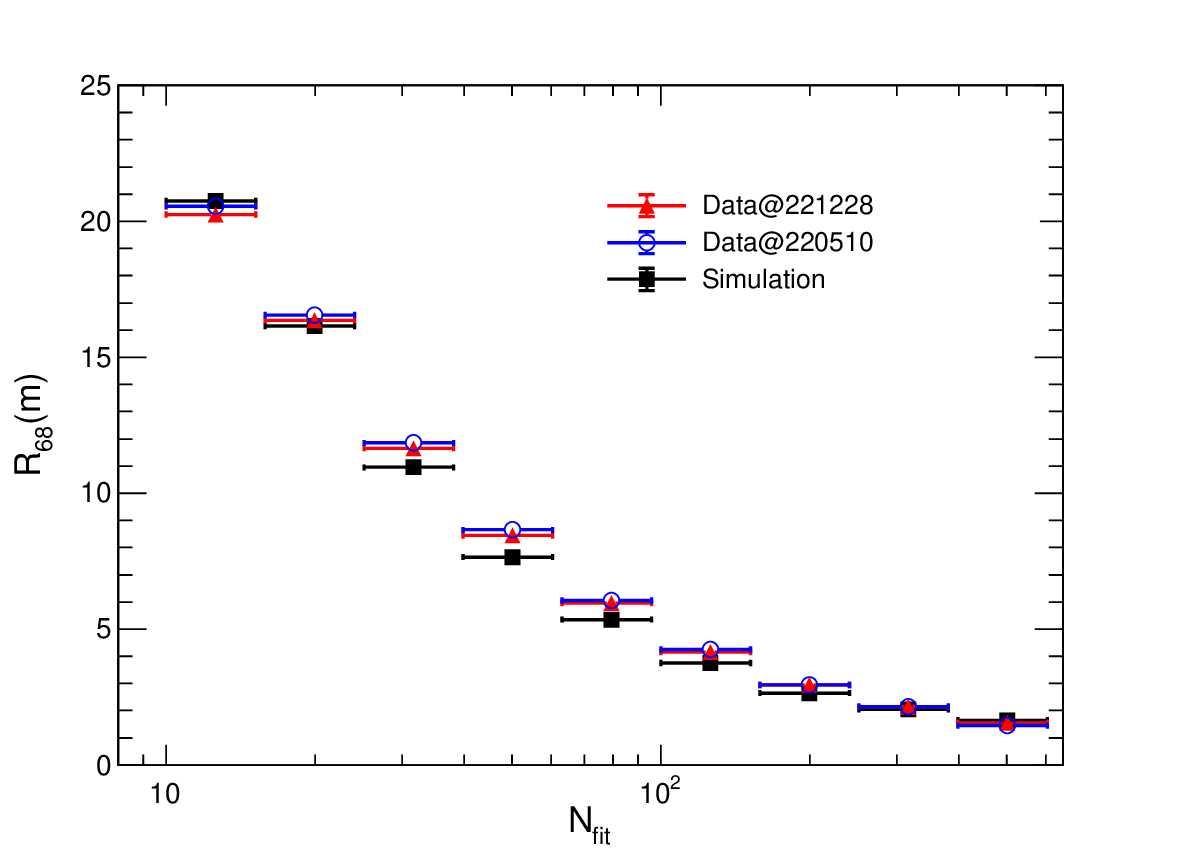}

\caption{ Angular resolution and core resolution as a function of the total N$_{\rm fit}$ (the number of fired EDs after filtering
out the noise)
  in the KM2A array by applying the odd-even method
to the experimental data and simulation data, respectively.}
\label{fig121}
\end{figure}

To directly compare the angular and core resolution between experiment and simulation,
the KM2A array is divided into two interspersed sub-arrays according to odd and even of the detector numbers, which is called the odd-even method \cite{oepaper19}.
 The two sub-arrays are used independently to reconstruct the direction and core of the same shower. Therefore, their difference divided by a factor of 2  provides an estimate of the resolution. The comparisons are presented in Fig. \ref{fig121}, which shows an agreement between experiment and simulation. The trends of the resolutions are consistent, but due to the fact that simulations cannot fully describe the inconsistency between detectors, and there are about 10\%-20\% difference between the model and real cosmic ray composition, the results of the simulation and the experiment still have insignificant differences. The comparison of the angular resolution between experiment and simulation has also been realized via observation of the gamma ray source Crab Nebula and the cosmic ray deficit from the Moon shadow \cite{aharon21}\cite{moonshadow}. The angular distributions of gamma ray events around Crab Nebula are fairly consistent with the point spread functions predicted by the simulation \cite{aharon21}. The angular resolutions for cosmic ray events obtained via the observation of the Moon shadow also shows a good agreement with the prediction of the simulation according to \cite{moonshadow}.

\section{Summary}
In summary, a simulation code for  the KM2A detector has been developed within the framework of the Geant4 toolkit, named as G4KM2A. This code is capable of simulating different stages of the array using a flexible geometry input method.
 To improve the running time of the code, some simplifications have been implemented, which can reduce the running time by at
least 30 times to meet practical requirements.
 In order to mitigate the problem of memory overflow, the simulation process flow has been adjusted by reading in the secondary particles of a shower  one by one in the particle generator. This method can be widely adopted in the simulations of similar large cosmic ray detectors  in further work.
Using the G4KM2A code, the response of the full KM2A array to cosmic rays has been simulated. The achieved trigger rate is roughly consistent with  the measurements, and the numbers of  electromagnetic particles  and  muons  yielded by the simulation are roughly consistent with the experimental data. The angular and core resolution also show a good agreement between experiment and simulation. These comparisons provide crucial evidence to  verify the reliability of the simulation code, which will be widely adopted for gamma ray astronomy and cosmic ray physical analysis in the future.

\vspace{3mm}
\bmhead{Acknowledgments}
We would like to thank all staff members who work at the LHAASO site above 4400 meters above sea level year-round to maintain the detector and keep the water recycling system, electricity power supply and other components of the experiment operating smoothly. We are grateful to Chengdu Management Committee of Tianfu New Area for the constant financial support for research with LHAASO data. We deeply appreciate the computing and data service support provided by the National High Energy Physics Data Center for the data analysis in this paper. This research work is also supported by the following grants: The National Key R\&D program of China under grants 2018YFA0404201, by the National Natural Science Foundation of China (NSFC) No.12022502, No.12205314, No. 12105301, No. 12261160362, No.12105294, No. U1931201. In Thailand, support was provided by the National Science and Technology Development Agency (NSTDA) and the National Research Council of Thailand (NRCT) under the High-Potential Research Team Grant Program (N42A650868).

\section*{Declarations}
\textbf{Conflict of interest} The authors declare that they have no conflict of interest.

\end{document}